\documentclass[acmsmall,screen]{acmart}

\usepackage{mdframed}
\usepackage{graphicx}
\usepackage{Templates/template}
\usepackage{multirow}
\usepackage{graphicx}
\usepackage{enumitem}
\usepackage{wrapfig}
\usepackage[capitalise]{cleveref}
\usepackage{array,multirow,graphicx}
\usepackage{hyperref}
\usepackage{hypcap}
\usepackage{xcolor}
\usepackage{amsmath}
\usepackage{booktabs} 
\usepackage{subcaption}
\usepackage{colortbl}
\usepackage{color}
\usepackage{listings}
\usepackage{balance}
\usepackage{colortbl}
\usepackage{tcolorbox}
\usepackage{url}
\usepackage{soul}
\usepackage{tcolorbox}
\usepackage{MnSymbol}
\usepackage{csquotes}

\usepackage{tikz}
\usetikzlibrary{arrows}
\usetikzlibrary{arrows.meta}
\usetikzlibrary{backgrounds}
\usetikzlibrary{calc}
\usetikzlibrary{positioning}
\usetikzlibrary{quotes}
\usetikzlibrary{scopes}
\usetikzlibrary{shapes}

\usepackage[font=small,labelfont=bf]{caption}
\usepackage{tcolorbox}
\definecolor{light-green}{rgb}{.5,1,.5}
\definecolor{light-pink}{rgb}{1,0.5,.5}

\floatstyle{ruled}
\newfloat{example}{thp}{lop}
\floatname{example}{Example }

\definecolor{codegreen}{rgb}{0,0.6,0}
\definecolor{codegray}{rgb}{0.5,0.5,0.5}
\definecolor{codepurple}{rgb}{0.58,0,0.82}
\definecolor{backcolour}{rgb}{0.95,0.95,0.92}

\definecolor{goals}{HTML}{28784D}
\newenvironment{standoutgoals}[1][]{%
  \mdfsetup{%
    frametitle={%
        \tikz[baseline=(current bounding box.east),outer sep=0pt]
        \node[anchor=east,rectangle,fill=goals]
        {\strut\footnotesize\textcolor{white}{Goal}};}}
  \mdfsetup{innertopmargin=4pt,linecolor=goals,%
    linewidth=2pt,topline=true,%
    frametitleaboveskip=\dimexpr-\ht\strutbox\relax,
    skipabove=4pt
  }
  \begin{mdframed}[]\relax%
    \label{goals:#1}}{\end{mdframed}}

\definecolor{rqs}{HTML}{aa5239}
\newenvironment{standoutrqs}[1][]{%
  \mdfsetup{%
    frametitle={%
        \tikz[baseline=(current bounding box.east),outer sep=0pt]
        \node[anchor=east,rectangle,fill=rqs]
        {\strut\footnotesize\textcolor{white}{#1}};}}
  \mdfsetup{innertopmargin=4pt,linecolor=rqs,%
    linewidth=2pt,topline=true,%
    frametitleaboveskip=\dimexpr-\ht\strutbox\relax,
    skipabove=4pt
  }
  \begin{mdframed}[]\relax%
    \label{rq:#1}}{\end{mdframed}}

\definecolor{findings}{HTML}{28784D}
\newenvironment{standoutfindings}[1][]{%
  \mdfsetup{%
    frametitle={%
        \tikz[baseline=(current bounding box.east),outer sep=0pt]
        \node[anchor=east,rectangle,fill=findings]
        {\strut\footnotesize\textcolor{white}{#1}};}}
  \mdfsetup{innertopmargin=4pt,linecolor=findings,%
    linewidth=2pt,topline=true,%
    frametitleaboveskip=\dimexpr-\ht\strutbox\relax,
    skipabove=4pt
  }
  \begin{mdframed}[]\relax%
    \label{finding:#1}}{\end{mdframed}}

\definecolor{promptcolor}{HTML}{2D4671}
\definecolor{promptcolor2}{HTML}{805A15}
\newenvironment{promptbox}[1]{%
  \mdfsetup{%
    frametitle={%
      \tikz[baseline=(current bounding box.east),outer sep=0pt]
      \node[anchor=east,rectangle,fill=promptcolor]
      {\strut\footnotesize\textcolor{white}{#1}};},
    innertopmargin=0pt,
    linecolor=promptcolor,
    linewidth=2pt,
    frametitleaboveskip=\dimexpr-\ht\strutbox\relax,
    skipabove=4pt
  }
  \begin{mdframed}[]\relax%
    \scriptsize\ttfamily 
    }{\end{mdframed}}

\definecolor{hlcolor}{HTML}{83C0A0}
\newcommand{\hlb}[1]{%
  \colorbox{hlcolor}{\{#1\}}%
}

\newenvironment{stackedprompttop}[1]{%
  \mdfsetup{%
    frametitle={%
      \tikz[baseline=(current bounding box.east),outer sep=0pt]
      \node[anchor=east,rectangle,fill=promptcolor]
      {\strut\footnotesize\textcolor{white}{#1}};},
    innertopmargin=0pt, 
    innerbottommargin=12pt, 
    linecolor=promptcolor,
    linewidth=2pt,
    frametitleaboveskip=\dimexpr-\ht\strutbox\relax,
    skipabove=0pt, 
    skipbelow=0pt, 
  }
  \begin{mdframed}[]\relax%
    \scriptsize\ttfamily 
    }{\end{mdframed}}

\newenvironment{stackedpromptmid}[2]{%
  \mdfsetup{%
    frametitle={%
      \tikz[baseline=(current bounding box.east),outer sep=0pt]
      \node[anchor=east,rectangle,fill=#2]
      {\strut\footnotesize\textcolor{white}{#1}};},
    innertopmargin=0pt, 
    innerbottommargin=4pt,
    linecolor=#2,
    linewidth=2pt,
    frametitleaboveskip=\dimexpr-\ht\strutbox\relax,
    skipabove=-1pt, 
    skipbelow=0pt, 
  }
  \vspace{-1em}
  \begin{mdframed}[]\relax%
    \scriptsize\ttfamily 
    }{\end{mdframed}}

\lstdefinelanguage{yml}{
  keywords={true,false,null,y,n},
  keywordstyle=\color{blue}\bfseries,
  sensitive=false,
  comment=[l]{\#},
  morecomment=[s]{/*}{*/},
  commentstyle=\color{purple}\ttfamily,
  stringstyle=\color{red}\ttfamily,
  morestring=[b]',
  morestring=[b]"
}

\newcommand\YAMLcolonstyle{\color{blue}\footnotesize}
\newcommand\YAMLkeystyle{\color{blue}\footnotesize}
\newcommand\YAMLvaluestyle{\color{black}\footnotesize}

\makeatletter

\newcommand\language@yaml{yaml}

\expandafter\expandafter\expandafter\lstdefinelanguage
\expandafter{\language@yaml}
{
  keywords={true,false,null,y,n},
  keywordstyle=\color{darkgray}\bfseries,
  basicstyle=\YAMLkeystyle,                                 
  sensitive=false,
  comment=[l]{\#},
  morecomment=[s]{/*}{*/},
  commentstyle=\color{purple},
  stringstyle=\YAMLvaluestyle,
  moredelim=[l][\color{orange}]{\&},
  moredelim=[l][\color{magenta}]{*},
  moredelim=**[il][\YAMLcolonstyle{:}\YAMLvaluestyle]{:},   
  morestring=[b]',
  morestring=[b]",
  literate =    {---}{{\ProcessThreeDashes}}3
                {|}{{\textcolor{red}\textbar}}1 
                {\ -\ }{{\mdseries\ -\ }}3,
}

\lst@AddToHook{EveryLine}{\ifx\lst@language\language@yaml\YAMLkeystyle\fi}
\makeatother

\newcommand\ProcessThreeDashes{\llap{\color{cyan}\mdseries-{-}-}}

\newcommand{\prompt}{Dev Prompt\xspace}
\newcommand{\prompts}{Dev Prompts\xspace}

\newcommand{\RQone}{How widespread are Bias and Bias-proneness in \prompts? How effectively can we  minimize these issues?}
\newcommand{\RQtwo}{How widespread is Vulnerability to injection attacks in \prompts? How effectively can we harden \prompts against them?}
\newcommand{\RQthree}{How widespread is sub-optimality of prompts in \prompts? How effectively can we optimize \prompts' performance?}
\newcommand{\PromptTool}{\texttt{PromptDoctor}\xspace}

\newcommand{\BiasedPromptsPercentage}{3.46\%\xspace}

\newcommand{\GenderBiasedPromptsPercentageFixed}{82.81\%\xspace}
\newcommand{\RaceBiasedPromptsPercentageFixed}{12.50\%\xspace}
\newcommand{\SexualityBiasedPromptsPercentageFixed}{50\%\xspace}
\newcommand{\BiasedPromptsPercentageFixed}{68.29\% \xspace}

\newcommand{\VulnerablePromptsPercentage}{10.75\%\xspace}
\newcommand{\VulnerablePromptsPercentageFixed}{41.81\%\xspace}

\newcommand{\NumberOfPromptsAttemptedToOptimize}{3,310\xspace}
\newcommand{\NumberOfPromptsOptimized}{37.1\%\xspace}

\definecolor{codeblue}{rgb}{0.26, 0.32, 0.45}
\definecolor{codegreen}{rgb}{0.22, 0.49, 0.32}
\definecolor{codepurple}{rgb}{0.44, 0.12, 0.49}
\definecolor{backgrey}{rgb}{0.96, 0.96, 0.96}

\lstdefinestyle{modernpython}{
    language=Python,
    backgroundcolor=\color{backgrey},
    basicstyle=\ttfamily\footnotesize\color{codeblue},
    keywordstyle=\color{codepurple}\bfseries,
    stringstyle=\color{codegreen},
    commentstyle=\itshape\color{gray},
    showstringspaces=false,
    frame=single,
    frameround=tttt,
    rulecolor=\color{gray},
    numbers=none,
    tabsize=4,
    xleftmargin=4pt,
    xrightmargin=4pt,
    breaklines=true,
    morekeywords={self, as, in, from, import},
}

\usepackage{codeanatomy}
\begin{document}
\renewcommand {\scriptsize} {\footnotesize}

\title{An Empirically-grounded tool for Automatic Prompt Linting and Repair: A Case Study on Bias, Vulnerability, and Optimization in Developer Prompts}


\author{Dhia Elhaq Rzig}
\affiliation{%
  \institution{University of Michigan-Dearborn}
  \city{Dearborn}
  \state{MI}
  \country{USA}
}
\email{dhiarzig@umich.edu}

\author{Dhruba Jyoti Paul}
\affiliation{%
  \institution{University of Wisconsin-Madison}
  \city{Madison}
  \state{WI}
  \country{USA}
}
\email{djpaul2@wisc.edu}

\author{Kaiser Pister}
\affiliation{%
  \institution{University of Wisconsin-Madison}
  \city{Madison}
  \state{WI}
  \country{USA}
}
\email{kaiser@pister.dev}

\author{Jordan Henkel}
\affiliation{%
  \institution{Microsoft (Gray Systems Lab)}
  \city{Madison}
  \state{WI}
  \country{USA}
}
\email{jordan.henkel@microsoft.com}

\author{Foyzul Hassan}
\affiliation{%
  \institution{University of Michigan-Dearborn}
  \city{Dearborn}
  \state{MI}
  \country{USA}
}
\email{foyzul@umich.edu}


\begin{abstract}

     The tidal wave of advancements in Large Language Models (LLMs) has led to their swift integration into application-level logic. Many software systems now use prompts to interact with these black-box models, combining natural language with dynamic values interpolated at runtime, to perform tasks ranging from sentiment analysis to question answering. Due to the programmatic and structured natural language aspects of these prompts, we refer to them as \textit{Developer Prompts}. Unlike traditional software artifacts, \prompts blend natural language instructions with artificial languages such as programming and markup languages, thus requiring specialized tools for analysis, distinct from classical software evaluation methods.

     In response to this need, we introduce \textit{PromptDoctor}, a tool explicitly designed to detect and correct issues of \prompts. \textit{PromptDoctor} identifies and addresses problems related to bias, vulnerability, and sub-optimal performance in \prompts, helping mitigate their possible harms. In our analysis of 2,173 \prompts, selected as a representative sample of 40,573 \prompts, we found that \BiasedPromptsPercentage contained one or more forms of bias, \VulnerablePromptsPercentage were vulnerable to prompt injection attacks. Additionally, \NumberOfPromptsAttemptedToOptimize were amenable to automated prompt optimization. To address these issues, we applied \textit{PromptDoctor} to the flawed \prompts we discovered. \PromptTool de-biased \BiasedPromptsPercentageFixed of the biased \prompts, hardened \VulnerablePromptsPercentageFixed of the vulnerable \prompts, and improved the performance of \NumberOfPromptsOptimized sub-optimal \prompts. Finally, we developed a \textit{PromptDoctor} VSCode extension, enabling developers to easily enhance \prompts in their existing development workflows. The data and source code for this work are available at~\cite{repl}.
\end{abstract}



\maketitle
\bibliographystyle{IEEEtran}

\keywords{Prompt Bias, Prompt Injection, Prompt Optimization}

\section{Introduction}
The rise of large language models (LLMs) has rapidly transformed modern software development, with these models becoming integral components in application logic~\cite{weber2024largelanguagemodelssoftware,HadiLargeLM}. Many systems now rely on structured prompts to interact with these black-box models by mixing natural language with dynamic runtime values. These structured prompts, termed Developer Prompts (\prompts), are fundamentally different from traditional software artifacts, requiring specialized analysis tools that go beyond classical prompt analysis methods~\cite{sourceprompt,pister2024promptset}. As \prompts combine natural language and programmatic elements, they introduce new challenges, such as biases, vulnerabilities, and sub-optimal performance that can impact overall system performance and reliability.

Existing research~\cite{guo2022auto,Clemmer2024,pryzant2023automatic,Wang_2023,thakur-2023} on prompts focused mainly on conversational prompts. Specifically, Guo et al.~\cite{guo2022auto} and Clemmer et al.~\cite{Clemmer2024} proposed techniques of reducing biased responses to conversational prompts by fine-tuning LLMs. Also noteworthy are the works of  Wang et al.~\cite{Wang_2023} and Pryzant et al.~\cite{pryzant2023automatic}, which focused on optimizing the performance of conversational prompts by hand-crafting and evaluating prompt engineering practices and applying computational modifications to LLMs, respectively. However, none of these works focused on \prompts: natural language prompts embedded in source code. \prompts represent a relatively new class of software artifacts, and their rapid proliferation introduces distinct challenges in software engineering. Unlike traditional code, \prompts are primarily composed of natural language, making them susceptible to a variety of issues, including bias, vulnerability to injection attacks, and performance sub-optimal performance. The rise of \prompts has even led to the emergence of a new role within software development: the "Prompt Engineer." Therefore, to address these recent and rapid changes in the software engineering landscape, our work adopts the following high-level goal:

\begin{standoutgoals}[Goal]
Develop tools and techniques to detect, analyze, and improve \prompts, addressing issues like bias~\cite{guo2022auto}, vulnerability~\cite{liu2024formalizing,yao2024promptcare}, and sub-optimal performance~\cite{pryzant2023automatic}, to support developers and prompt engineers in creating more reliable interactions with Large Language Models.
\end{standoutgoals}
\vspace{0.5em}

\prompts introduce challenges distinct from traditional code due to the vagueness and ambiguity inherent in natural language. This makes them prone to bias~\cite{guo2022auto}, injection attacks~\cite{liu2024formalizing}, and sub-optimal performance~\cite{shi2023don}. Additionally, the integration of natural language with traditional code presents a largely uncharted area for automated analysis. In this work, we examine \prompts in their operational contexts, focusing on how dynamic value interpolation and runtime factors influence their reliability, building on the dataset of \prompts collected from PromptSet~\cite{pister2024promptset}. We now focus on three primary challenges with \prompts---bias, vulnerability, and performance---each posing a significant risk in software development. Let’s examine them in more detail:

\begin{enumerate}[topsep=1pt, itemsep=0pt, partopsep=0pt, parsep=0pt, labelindent=1pt,leftmargin=17pt]
    \item \textbf{Bias}: \prompts must be carefully crafted to avoid both explicit and implicit biases, as even subtle wording can have a substantial impact. One common prompt design strategy is to provide the model with a “persona” to better perform a task. However, it is easy for biases to be unintentionally encoded into these personas, which can influence the model’s behavior in undesirable ways, such as propagating bias and potentially creating societal harm.

    \item \textbf{Vulnerability}: Dynamic variable interpolation in \prompts makes them vulnerable to prompt injection attacks. Understanding how to properly sanitize user input before integrating it into prompts is an ongoing challenge. Without careful input validation, many software systems risk exposing too much control to users, making them susceptible to malicious prompt manipulation that may lead to a leak of sensitive information, for example, thus potentially causing  customer harm.

    \item \textbf{Sub-optimal Performance}: Crafting effective prompts is often seen as more of an art than a science, which means many existing \prompts are likely under-performing and can benefit from optimization. Performance in this context is task-specific and not directly related to the software's runtime but rather to the accuracy and relevance of the model's output for the given task. Relying on sub-optimal prompts can cause customer users to lose trust in the products that rely on them, potentially causing product reputation harm.
\end{enumerate}

To highlight these issues, we provide an illustrative example in~\autoref{fig:source-prompt-example} from the GitHub project \textit{blob42/Instrukt}. At first glance, the prompt seems innocuous, yet it suffers from all three challenges outlined above. By using the typically female name ``Vivian,'' the prompt risks biasing the model toward generating stereotypically female-coded responses. Empirically, we also found that this prompt is vulnerable to prompt injection attacks via the variable \texttt{context} and that using strong imperative commands, such as "you MUST ..." leads to a \textasciitilde 20\% gain in prompt adherence on a synthetic dataset designed for this prompt. In any case, ~\autoref{fig:source-prompt-example} shows just how easy it is for \prompts to have many issues---issues that require new tools and new research to identify and fix.

\begin{figure}[h]
\begin{promptbox}{Example of a Flawed Prompt}
You are Pr. Vivian. Your style is conversational, and you always aim to get straight to the point. Use the following pieces of context to answer the users question. If you don't know the answer, just say that you don't know, don't try to make up an answer. Format the answers in a structured way using markdown. Include snippets from the context to illustrate your points. Always answer from the perspective of being Pr. Vivian. \\
\verb|----------------| \\
\hlb{context}
\end{promptbox}
\caption{Example of a prompt with bias and injection issues from GitHub project: \textit{blob42/Instrukt}}
\label{fig:source-prompt-example}
\end{figure}

Within this research work,
we analyzed how widespread these three issues are in Open Source Software (OSS) \prompts, and we offer an easy-to-use solution for mitigating them. Using empirically validated LLM-powered processes, we perform a large-scale analysis on PromptSet~\cite{pister2024promptset}, a collection of \prompts mined from open-source projects, to detect the prevalence of bias and injection vulnerability. We also perform an in-depth analysis to examine a varied set of \prompts of different task categories to illustrate the prevalence of sub-optimality. Then, we propose \PromptTool, a bespoke solution to address these issues within \prompts. \PromptTool utilizes automatic issue discovery and prompt rewriting processes that build on a generation-evaluation paradigm to automatically correct flaws within \prompts, and is both technology and LLM independent. Indeed, \PromptTool is easily integrated in development processes, allowing developers to detect and correct the aforementioned issues in \prompts, all while avoid the expensive process of Model Fine-Tuning altogether. \PromptTool analyzes \prompts before their deployment and automatically mitigates any issues they may contain, thus preemptively avoiding any potential harms these issues can create. Finally, we evaluate \PromptTool and report its results when used on the flawed \prompts we identified. We achieve these goals by answering the following research questions:

\begin{itemize}[topsep=1pt, itemsep=0pt, partopsep=0pt, parsep=1pt, labelindent=1pt,leftmargin=10pt]
    \item \textbf{RQ1:} \RQone
    \item \textbf{RQ2:} \RQtwo
    \item \textbf{RQ3:} \RQthree
\end{itemize}

To answer these questions, we design and evaluate various LLM-powered processes that form the foundation of \PromptTool.
\PromptTool allowed us to establish that \BiasedPromptsPercentage of \prompts are prone to generating biased responses, and that \VulnerablePromptsPercentage  are vulnerable to prompt-injection attacks. We also found that 36\% of Question-Answering \prompts quantitatively under-performed when tested against real-world benchmarks. 

Using \PromptTool, we were able to improve a portion of these flawed \prompts by de-biasing \BiasedPromptsPercentageFixed of the biased \prompts, hardening \VulnerablePromptsPercentageFixed of vulnerable \prompts, and optimizing \NumberOfPromptsOptimized of the sub-optimal \prompts.

Our contributions are:
\begin{itemize} [topsep=1pt, itemsep=0pt, partopsep=0pt, parsep=1pt, labelindent=1pt,leftmargin=10pt]
    \item The first large-scale analysis of OSS \prompts, uncovering \textbf{bias}, \textbf{vulnerability}, and \textbf{sub-optimality} in them.
    
    \item Creation and Evaluation of \PromptTool, a novel solution to detect and fix \textbf{bias}, \textbf{vulnerability}, and \textbf{sub-optimality} in \prompts, made available as a VS Code extension.
    
    \item Our empirical findings and observations on \prompts have uncovered a new area that could inspire and guide future research directions.
    
\end{itemize}


\section{Background}
\label{sec:background}

\subsection{Large Language Models and \prompts}
\label{sub:sec:language-models-prompts-bg}
Large Language Models (LLMs) have emerged as a transformative advancement in natural language processing~\cite{Bharathi_2024,Makridakis_2023}. 
Unlike earlier models, LLMs can scale to billions of parameters and vast volumes of training data~\cite{OpenAI_2024,Dubey_2024}, which
has led to emergent capabilities, such as in-context learning~\cite{Schick_2021,Dang_2022},
The output of a language model is guided by the input context, or prompt. Prompts can take various forms: questions, statements, multi-turn dialogues, etc. 
However, in this work, we focus on prompts written by developers and used within software applications, which we refer to as \textit{Developer Prompts}~\cite{pister2024promptset} or \textit{\prompts} for short. These prompts operate in more constrained contexts, typically embedded within specific methods to generate a targeted output. 
\prompts are generally not directly visible to or editable by the user; instead, the user can only influence them indirectly by setting variables or parameters that are interpolated into the \prompt. A notable exception is prompt playgrounds or model comparison tools, which may allow users to specify the entire prompt. In addition, \prompts are usually used in one-off interactions with the model rather than being part of a multi-turn dialogue. Furthermore, the context in which \prompts are used is often domain-specific, such as text summarization or translation based on user input.
An example of a \prompt in source code is shown in ~\autoref{lst:source_prompt}.
The proliferation of LLMs 
has driven their integration into traditional software systems via \prompts, offering impressive capabilities but also introducing new challenges. 
Addressing the challenges posed by these ``hybrid'' software systems---those combining traditional software logic with LLM-powered components---will require the development of new tools and strategies.


\begin{lstlisting}[style=modernpython, caption=Example of a \prompt from \texttt{ownsupernoob2/Blimp-Academy-Flask},label=lst:source_prompt]
def product_observation(prompt_product_desc):
    response = openai.Completion.create(
        model="text-davinci-002",
        prompt="The following is a conversation with an AI Customer Segment Recommender.... 
        AI, please state a insightful observation about " + prompt_product_desc + ".",
        temperature=0.9, max_tokens=...)
    return response['choices'][0]['text']
\end{lstlisting}


\subsection{Bias in Language Models}
\label{sub:sec:bias-in-lms-bg}

Like other Machine Learning models, LLMs can learn biases from the data they are trained or fine-tuned on. These biases can have a cascading effect on the software that uses these models. For example, a bank loaning software that used an ML model to determine creditworthiness was found to be biased against people of certain intersectional groups~\cite{Kim_2023}, even though protected attributes such as race and gender was not exposed as to the model.
In the context of LLMs, bias and potential for bias stubbornly persist~\cite{Breckenridge_2024,Gallegos_2024,Cheng_2023}. Furthermore, LLMs risk further propagating these biases and stereotypes~\cite{Omiye_2023,Hofmann_2024}, in ways that can be hard to directly see or detect beforehand. For example, as discussed by Cheng et al.~\cite{Cheng_2023}, LLMs are more likely to make assumptions about people of a certain gender and race, even when the prompts fed into the model do not contain any information that would lead to these assumptions. These assumptions and biases not only risk generating biased responses, but also risk causing harm to the people who interact with the software that uses these models, by altering their internal behavior and decision-making processes, all while providing little to no transparency about their reasoning to the end-user, such as the case of the creditworthiness model. Hence, it is imperative to detect and fix bias in \prompts to avoid causing Societal harm. 

\subsection{Vulnerability in Language Models}
\label{sub:sec:vulnerability-in-lms-bg}

\prompts present a new attack vector in software applications. Prompt-injection~\cite{Rossi_2024} is a novel technique where attackers inject specific phrases into a prompt to cause the LLM to behave in ways against its design intentions as set by the software's developers. Prompt-injection attacks can trigger failures such as unwanted responses that misuse company's resources, such as misusing a company customer support bot to get free access to a paid LLM service. In more serious scenarios, it can lead to the exposure of sensitive information, where it can coax the LLM to reveal confidential information about the company's inner machinations, such as private email addresses~\cite{White_2023}, or even the physical location of company resources. These attacks can also coax intellectual property contained within the prompt by sharing its internal text, which makes further misuse even easier~\cite{Hui_2024}. The pace of the development of Prompt-injection attacks~\cite{Rossi_2024,Liu_2023,Liu_2024,Yu_2023,Greshake_2023} is indicative of the growing threat they pose to software applications that use LLMs. Thus, it is important to detect and fix vulnerabilities in \prompts to avoid causing Customer harm.

\subsection{Performance of Language Models}
\label{sub:sec:performance-of-lms-bg}
%
While LLMs can surprisingly well in certain tasks, such as text generation and text summarization, their performance remains mediocre at tasks such as math and logic~\cite{Sawada_2023,Sebler_2024}. Further complicating matters, there is no standardized automatic way to evaluate or optimize Prompts. Prompt engineering is an emerging field that attempts to address these shortcomings by providing a systematic way to write prompts~\cite{White_2023_Pattern,Amatriain_2024}. There are many competing practices such as Chain-of-Thought (COT), where the prompt asks the LLM to explain its reasoning step by step; and ``N-Shot'' Prompting, where the prompt contains example input-output pairs, and instruction alignment where the prompt contains rules for the model to follow. However, prompt engineering remains a largely manual task with no quantitative guidelines to follow. Thus, there is a need for the automatic evaluation and optimization of \prompts to ensure good performance of the systems that rely on them, and avoid Product-Reputation harm. 

\section{Research Approach}

\label{sec:approach}
\begin{figure*}[htbp]
    \centering
    \includegraphics[width=\linewidth]{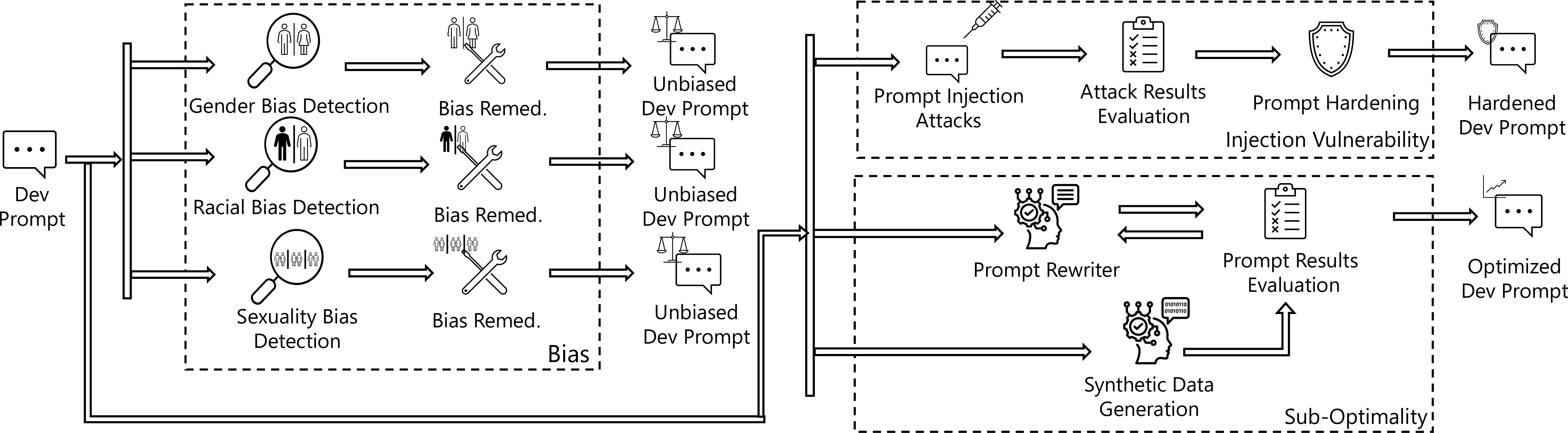}
    \caption{Overview of the Research Approach}
    \label{fig:overall}
\end{figure*}


\subsection{Data Preparation}
\label{sub:sec:data-preparation}

\subsubsection{Dataset Selection}

As discussed in~\autoref{sub:sec:language-models-prompts-bg}, we focus within this work on \prompts sourced from PromptSet~\cite{pister2024promptset} which contains 61,448 unique \prompts collected from 20,598 OSS projects. However, no examination the quality of these prompts or significant cleaning operations were performed during PromptSet's creation. Indeed, upon manual inspection, we found that it contained a number of toy prompts that are not representative of "production-quality" \prompts. We tackle cleaning this dataset in~\autoref{sub:sub:sec:dataset-cleaning}.

In addition, \prompts are not just static strings of text: they interweave structured natural language with traditional software languages. Many contain variables that are added at interpolated before being sent to the LLM. Since \prompts are commonly represented as variable(s) in source code, they do not have a standardized representation. Values might be interpolated or concatenated dynamically depending on on the control flow of the program. For example a \prompt might take the form of \texttt{"this is \{x\}."} or \texttt{"this is "+x+"."}, and these representations are stored exactly as they are extracted from source code in PromptSet. We refer to the variables in \prompts as \textit{Prompt Holes}, and as their values are generally defined at runtime via user input, it's difficult to determine what they should be within the context of the different analyses we perform within this work. We attempt to address both of these issues via the processes detailed in~\autoref{sub:sub:sec:prompt-parsing}.


\subsubsection{Dataset cleaning}
\label{sub:sub:sec:dataset-cleaning}


While PromptSet contains a large and diverse set \prompts, it was important to perform some data cleaning processes to maximize high quality prompts in our experimental set. We identified that 25\% of the \prompts PromptSet contained had a length of 31 characters or less. Upon manual inspection, we found most were out of distorted or helper-prompts of little significance, hence we eliminated them from the set of prompts we analyzed. 45,747 \prompts remained. We also removed  non-English prompts, further eliminating 5,174 (11.31\%) prompts which contain non-ASCII and non-Emoji characters. Future research could focus on these \prompts to determine if the approaches proposed within this work can be applied to prompts in other languages as well. 40,573 prompts remained after this process.

\subsubsection{Prompt Parsing}
\label{sub:sub:sec:prompt-parsing}

\begin{figure}[h]
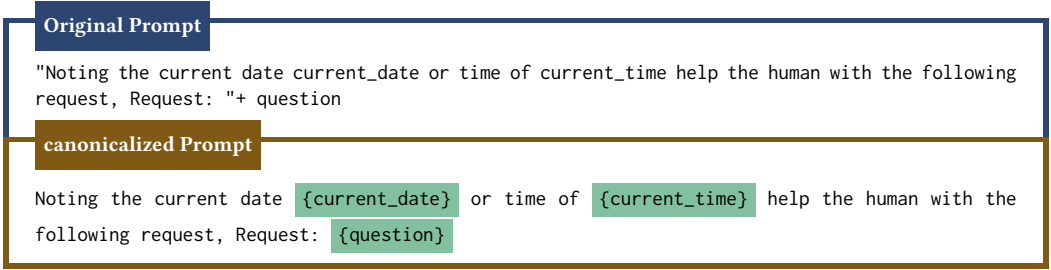

\begin{stackedprompttop}{Original Prompt}
"Noting the current date {current\_date} or time of {current\_time} help the human with the following request, Request: "+ question
\end{stackedprompttop}
\begin{stackedpromptmid}{canonicalized Prompt}{promptcolor2}
Noting the current date \hlb{current\_date} or time of \hlb{current\_time} help the human with the following request, Request: \hlb{question}
\end{stackedpromptmid}
\caption{A prompt from \texttt{zekis/bot\_journal}, before and after canonicalization}
\label{fig:canonicalization-example}
\end{figure}

\paragraph{Prompt Canonicalization}
\label{sub:sub:sec:canonicalization}
Via this processes, we standardized the presentation of the different \prompts into a universal, canonical, representation that was easier to programmatically parse and process. This processes relies on static parsing and regex matching to determine the different variable holes within a \prompt. The resulting canonical representations preserves the original \prompt's text along with the holes inter-weaved within. The prompt holes are delineated within special characters (\{ and \}). An example of the process of canonicalization is shown in~\autoref{fig:canonicalization-example}.


\paragraph{Prompt Patching}
\label{sub:sub:sec:patching}
After standardizing the representation of \prompts, we set out to create appropriate mock values for their different holes in order to ground our consecutive analyses in realistic usage scenarios of Dev Prompts. Due to their aptitude in generative tasks, we utilize LLMs to generate these values. However, the scope from which we can determine the appropriate values for the different prompt holes is not immediately clear, as the containing method or class may not always contain informative comments or names, and a ReadMe file may be too high-level to be relevant for the value of a single Prompt Hole.
Context window limitations further complicating matters, as a single class, method, or file may exceed the window size in some cases.
Hence, in order to generate mock values for these variables, a process we refer to as \textit{Patching} the prompt, we rely on the \prompt's text and the variable name corresponding to the Prompt Hole, thus making this process localized and easily transferable to other contexts. To this end, we hand-crafted a prompt following the practices discussed by Sahoo et al.~\cite{Sahoo_2024}, and send it to the LLM to generate mock values for each Prompt Hole. This handcrafted prompt is given at~\cite{repl}.

In the case of a prompt containing multiple holes, there were two possible approaches: either patching the holes in parallel in an independent manner or patching them sequentially in a dependent manner. In parallel patching, the mock value for each hole is generated independently of the other prompt hole values, which is formalized in~\autoref{eq:parallel-patching}. In sequential patching, the mock value generated for prompt hole x is dependent on the values generated for all the prompt holes that appear before it in the \prompt, this is formalized in~\autoref{eq:parallel-patching}.
We validated that generating values for the variables sequentially in their order of appearance, 
patched prompt holes in a more consistent and logical fashion than via parallel generation. 

\begin{equation}
    \scriptstyle 
    \label{eq:sequential-patching}
 val(h_x|patch(prompt, [val(h_x-1), val(h_x-2), ... val(h_1)]))
\end{equation}
\begin{equation}
    \scriptstyle 
    \label{eq:parallel-patching}
     val(h_x|prompt),val(h_x-1|prompt), val(h_x-2|prompt), ... val(h_1|prompt)
\end{equation}

For the sake of simplicity and cost efficiency, during bias and vulnerability detection and remediation, we restrict our patching process to generally only one value for each Prompt Hole, as generating multiple values can lead to an exponential number of combinations possible. An example of a canonicalized prompt and its corresponding generated values via the process of patching is shown in~\autoref{fig:patching-example}. Furthermore, we add multiple prompt-level and code-level mechanisms to ensure that this process does not introduce bias or bias-proneness in prompts.



\begin{figure}[h]
\begin{stackedprompttop}{canonicalized Prompt}
Noting the current date \hlb{PLACEHOLDER\_1} or time of \hlb{PLACEHOLDER\_2} help the human with the following request, Request: \hlb{PLACEHOLDER\_3}
\end{stackedprompttop}
\begin{stackedpromptmid}{Generated Values}{promptcolor2}
\hlb{PLACEHOLDER\_1} --> today \\
\hlb{PLACEHOLDER\_2} --> 3:00 PM \\
\hlb{PLACEHOLDER\_3} --> what are my upcoming meetings for the rest of the day?
\end{stackedpromptmid}
\caption{Example of Prompt Patching with \prompt taken from \texttt{zekis/bot\_journal}}
\label{fig:patching-example}
\end{figure}

\paragraph{Optimization Dataset Synthesis}
For the purpose of \prompt optimization, we extend the patching process to create synthetic datasets, which serve as input values assigned to the different prompt holes of the \prompt we aim to optimize. A synthetic dataset is comprised of multiple values for each Prompt Hole, values that are generated by the patching process are optimized for creativity and diversity to mitigate duplication as the quantity of patches per \prompt increases. In practice, these patches are generated in a sequential manner, where we use stratified temperatures during patch generation and include a few randomly-selected previously-generated patches for the same hole as examples to avoid duplication during the generation process. Within our hand-crafted prompt also includes guidelines as a context and we enforce a strict response format to ensure the data generated will conform to the original \prompt's structure~\cite{li2024synthetic}.

\subsubsection{Dataset Sampling}

After applying the cleaning process in~\autoref{sub:sub:sec:dataset-cleaning}, we obtained a set of 40,573 \prompts. The size of the set was still too computationally and financially expensive to use for all our analyses, especially as each technique relies on multiple LLM calls per \prompt. Hence, we set out on selecting a representative sample this larger set. After applying the canonicalization process from~\autoref{sub:sub:sec:canonicalization}, we extracted the number of prompt holes within the different \prompts. We found that 20,620; 9,427; 6,154; 2,204; 1,503; 464; and 651 \prompts had respectively 0; 1; 2; 3; 4; 5; and 6+ prompt holes. 
We grouped \prompts with more than 5 holes into one large group after finding a significant drop in the numbers for \prompts with exactly 6 holes or exactly 7 holes in comparison to \prompts with exactly 5 holes, where they were 63.44\% and 82.02\% smaller, respectively. 





From these results, prompt holes emerged as an appropriate  stratification criteria to help guide our process of selecting a representative population of \prompts. Hence, we applied the corresponding  random sampling from each strata with a 95\% confidence - 5\% error. After applying this process, we created a set comprised of  378, 370, 362, 328, 282, 211, and 242 \prompts with respectively 0, 1, 2, 3, 4, 5, and 6+ prompt holes, 
which we use for the rest of our analyses.

\paragraph{Optimization Dataset}
When optimizing \prompts, we shift our focus to categorizing the \prompt into a specific task categories. These four task categories are: question \& answer, grammar correction, summarization, and translation. We refert to the grammar correction, summarization and translation prompts as Grounded task prompts. Question \& answer (QA) prompts are open ended requests of a language model, such as the one shown in~\autoref{fig:source-prompt-example}. We filter PromptSet for short, English-language \prompts between 20 and 200 characters long. Additionally we use a mood filter from SpaCy to select only imperative or interrogative prompts \cite{Honnibal_spaCy_Industrial-strength_Natural_2020}. This process extracted 3,310 QA prompts for us to optimize. From these, we randomly sample 100 to conduct experiments.

For grammar correction, we filter by searching for keywords related to grammar and punctuation. Then, we augment this set by adding \prompts which are semantically similar to the \prompts which had keyword matches. This resulted in a dataset of 36 grammar correction \prompts. Using similar techniques, We also find a handful of 7 translation and 4 summarization \prompts.

\subsection{Addressing Bias}
\label{sub:sec:bias}
\subsubsection{Bias Detection}
\label{sub:sub:sec:bias-detection}

The issue of detecting biases within Natural language text remains a fraught and complicated issue. Biases can come in many shapes and forms, and can stem from multiple factors. The approach we designed to detect biases is generic, and  we use it to focus on three Biases that have been documented extensively within existing literature around Bias in software: Gender-Bias~\cite{Misa2022GenderBI, algs_Sexist, Vorvoreanu_2019, Guzma_2024,Terrell_2017}, Race-Bias~\cite{Kim_2023,Bogen_2019,Sap_2019,Obermeyer_2019}, and Sexuality-Bias~\cite{Maji_2024,Sultana_2021,Felkner_2023}. We believe our approach can easily be extended to detect other types of biases as well, but we leave this exploration to future research works.

Our approach is simple but effective, as discussed within~\autoref{sec:background}, LLMs are a great tool for linguistic and textual analyses, hence we leverage them via a hand-crafted prompt, available in our replication package~\cite{repl}, that specifies the type of bias we're attempting to detect
along with a patched version of the prompt we're evaluating. This prompt generates a JSON file, making it easy to programmatically ingest its results. This JSON contains three fields: whether the \prompt is Explicitly Biased, whether the \prompt is prone to generating biased responses, and an explanation behind the evaluation given. We distinguish between explicit bias and bias-proneness since, as documented by previous research~\cite{Cheng_2023}, \prompts without explicit bias can still cause LLMs to generate biased-responses. 


Following the recommendations of Radford et al.~\cite{Radford_2019} and Brown et al.~\cite{brown_2020}, we designed a Zero-Shot, One-Shot, and Multi-Shot, and  measured their performance against established benchmarks for the bias-detection as detailed in~\autoref{sub:sub:sec:bias-benchmarking}, and found that the Multi-shot version performed best. Hence, each bias-detection prompt also includes three example inputs: one explicitly-biased, one bias-prone, and one non-biased example. Each example was accompanied with an example expected JSON response. For added transparency, \PromptTool reports to users the results of this evaluation along process including the explanation behind the evaluation.

\subsubsection{Bias Remediation}
\label{sub:sub:sec:bias-remediation}
While there are many recommendations regarding methods to rewrite prompts to improve their performance~\cite{Yang_2024,Fernando_2023,Agarwal_2024,pryzant2023automatic}, there is no established generic method to rewrite prompts to de-Bias them. Hence, we created an automatic de-Biasing method within \PromptTool, that relies on a prompt generation-evaluation loop. Due to the prowess of LLMs when in text-generation, we also rely on their assistance during this process.

First, we evaluate the original prompt given by the developer via the method detailed in~\autoref{sub:sub:sec:bias-detection}.
Second, if the prompt is determined to be biased or bias-prone, we generate 5 rewrites of the developer-prompt with the goal of minimizing bias and bias-proneness, via a hand-crafted prompt, available in our replication package~\cite{repl}.
Third, we evaluate each of these rewrites for bias. If we determine some of them is also biased or bias-prone, we isolate them and run the same generation-evaluation loop on each of them. 
We stop this process when we have at least five new non-biased and non-bias-prone \prompts. We then supply these rewrites sorted by distance\footnote{Distance $=$ Number of iteration that generated the new \prompt} to the developer. We give developers multiple variants to promote higher flexibility for improved tool adoption~\cite{Dang_2023}. We limit this process for 10 iterations in order to avoid running indefinitely and generating \prompts that are too different from the original \prompts.

\subsection{Addressing Injection Vulnerability}
\label{sub:sub:sec:vulnerability}
\subsubsection{Vulnerability Detection}
\label{sub:sub:sec:vulnerability-detection}
There is a variety of Prompt Injection attacks possible when interacting with LLMs~\cite{Rossi_2024,Liu_2023,Liu_2024,Yu_2023,Greshake_2023}, however, as discussed within~\autoref{sub:sec:language-models-prompts-bg}, interacting with them via \prompts is generally more constrained than other scenarios. \prompts deployed within a project are generally not directly accessible or editable, and are sent to an LLM after variables within these prompts are interpolated via values shaped by by users. Hence, we focus on injection attacks that are deployed via inserting a malicious string within specific location(s) of the prompt.

Testing a prompt for injection vulnerability relies on a collection of 42 known attacks from a corporate dataset\footnote{Double blind policy forbids from being specific.} and the open web. Each of these attacks are formulated to produce a specific un-common target string as expected results. For each \prompt we examined, we first perform the canonicalization process discussed in~\autoref{sub:sub:sec:canonicalization}, and then for each Prompt Hole, we inject a specific attack into it and patch the remaining holes via the prompt patching process, detailed in~\autoref{sub:sub:sec:patching}. We then send the \prompt to the LLM, and inspect the corresponding response to determine whether the attack was successful by scanning for the expected target responses. We repeat this process for all the attacks we have in our collection and all of the prompt holes within the prompt. Since these attacks were independent of each other, we performed this process in a parallelized manner, and the equation in~\autoref{eq:parallel-injection} shows a formalization of this process. It's notable that \PromptTool reports which holes that served as successful injection points for these attacks to end-users.

\begin{equation}
 \scriptstyle 
    \label{eq:parallel-injection}
    Vulnerability_{prompt} = \sum_{i=1}^{n}attack_i(h_{pos},\mathop{\sum_{j=1}^{m}}_{j\neq pos} patch(prompt, hole_{j}))
\end{equation}

\subsubsection{Vulnerability Remediation}
\label{sub:sub:sec:vulnerability-remediation}
Similar to~\autoref{sub:sub:sec:bias-remediation}, there is no established way to harden`' prompts against prompt-injection. Instead, most existing methods focus on Model Fine-Tuning~\cite{Suo_2024,Liu_2023,Yi_2024} to defend against to these attacks. Furthermore, some systems add other layers of security to preemptively detect attack strings or compromised responses and then respond to those attacks with a generic denial response~\cite{PatrickFarley_2024}. As discussed in~\autoref{sec:background}, these approaches come with their own costs and caveats, hence, we implement a new Prompt hardening process within \PromptTool as a less-expensive and easier-to-deploy complement to these approaches. Similar to~\autoref{sub:sub:sec:bias-remediation}, this process also relies on a generation-evaluation loop. 
First, we analyze a \prompt to determine if it's vulnerable as described in~\autoref{sub:sub:sec:vulnerability-detection}.
Second, if the \prompt is vulnerable, we ask the LLM via a hand-crafted prompt, given at~\cite{repl}, to generate five new prompts that are a rewrite of the original prompt, and which are hardened against a specific attack. 
Third, for each generated \prompt, we verify that it contains the same prompt holes as the original \prompt, and then evaluate it for vulnerability it as described in~\autoref{sub:sub:sec:vulnerability-detection}. 
Finally, if  all of the attacks in our set fail against one of these new \prompts, we consider this \prompt hardened, else, we add it to our list of vulnerable \prompts for future generation-evaluation loop executions.
Empirically, we found that generating a hardened \prompt requires multiple iterations and an associated high cost, hence, we limit this process to 10 iterations, and we stop when we have one new hardened \prompt. Furthermore, we found that \prompts with the lowest number of vulnerable holes were more likely to be successfully hardened, hence we sort the vulnerable \prompts by the number of vulnerable holes before starting a new iteration of the hardening process

\subsection{Addressing Sub-Optimality}
\label{sub:sec:optimization}
\subsubsection{Sub-optimality Detection}
By default, we assume all \prompts are sub-optimal and  present an opportunity for possible automatic improvement, especially since previous research~\cite{Zamfirescu_2023} found that many non-experts struggle with prompting. Indeed, even a prompt writing expert cannot empirically prove their design has reached an optimal state. In addition, any evaluation processes may be hindered by the lack of a dataset.
Therefore, we design \PromptTool that automatically optimizes a \prompt against a synthetic dataset to improve performance without requiring costly manual exploration or data collection by the developer.

\subsubsection{Optimization Process}
The \PromptTool optimization process is summarized as follows: 1. Generation of synthetic training and test datasets based on the \prompt, 2. Creation of a few seed \prompts based on good prompting strategies written by OpenAI~\cite{OpenAIPE} and Anthropic~\cite{AnthropicPE}, 3. Evaluation of all the generated \prompts on the synthetic data, 4. Execution of a self improving optimization algorithm to discover new \prompts as described in~\autoref{eq:optimization}, 5. Repetition  of steps 3-4 until no performance gain is observed on the training dataset.

\paragraph{Seed \prompts Generation}

Self-optimization strategies are sensitive to their starting seed~\cite{Yang_2024}. Hence, to mitigate falling into local minima, we propose multiple strategies to diversify the initial seed configuration for optimization. 
Primarily, we generate seed \prompts based on prompt rewriting suggestions provided by OpenAI and Anthropic. The generative meta-prompt receives\ a sample of five different \prompt principles, such as "Use Delimiters" or "Add to your \prompt the following phrase: `Ensure your answer is unbiased and does not rely on stereotypes'", out of 26 principles detailed at~\cite{repl}. Additionally, we vary the generative temperature to induce both precise and creative responses. We validate that the resulting generated \prompts are well formatted and has the same prompt holes as the initial \prompt.

\paragraph{Optimization}

We formalize the optimization algorithm as follows:

\begin{equation}
    \scriptsize
    \label{eq:optimization}
    Opt_{i} = M_{1}(t_{1}, s_{1}, s_{2}, ..., s_{n});~~~s_{j} = \sum_{k}f(., M_{2}(t_{2}, p_{j},D_{k}))/K
\end{equation}

where $i$ is the step count, $M$ a language model, $t$ a meta-prompt, and $s_{j}$ the tuple of a \prompt $p_j$ and its average score received across a dataset $D$ of size $K$. The $i=0$ step is a special case, where $p_{j}$ is sourced from the seed \prompts. Subsequent steps utilize the highest scoring $n$ \prompts generated across all steps of the algorithm. Finally, $f$ represents a task-based evaluator as described below. $M_1$ specifically is used to generate new candidate \prompts, $M_2$ generates responses based on candidate \prompts and elements from the synthetic training dataset.

\paragraph{Evaluation}
After the seed \prompts are generated and after each optimization step, we evaluate the quality of newly generated \prompts on the synthetic training dataset. We select a scoring criteria based on the categorization of the initial \prompt.

    \noindent\textbf{Translation.} We use the BLEU metric to compare the quality of the generated translation with the reference synthetic translation~\cite{bleumetric}. 
    \begin{equation}
        \scriptsize
        s_{j} = \sum_{k}BLEU(D_{k}["translation"],M_{2}(p_{j}, D_{k}["source"]))
    \end{equation}
    \noindent\textbf{Summarization.} We compare the semantic similarity of the generated summary with the reference summary using the cosine similarity of the embedded representations~\cite{reimers-2019-sentence-bert}. Using embedding function $E$:
    \begin{equation}
        \scriptsize
        s_{j} = \sum_{k}cos_sim(E(D_{k})["summary"],E(M_{2}(p_{j}, D_{k}["source"])))
    \end{equation}
    \noindent \textbf{Error Correction.} We use the GLEU metric to compare the quality of the generated phrase with the reference synthetic phrase~\cite{mutton-etal-2007-gleu}.
    \begin{equation}
        \scriptsize
        s_{j} = \sum_{k}GLEU(D_{k}["correct"],M_{2}(p_{j}, D_{k}["source"]))
    \end{equation}
    \noindent \textbf{QA Refinement.} We utilize the ``LLM as a judge'' technique to score QA tasks due to their highly varied nature~\cite{Zheng2024}. As a preprocessing step, each QA \prompt generates a corresponding scoring prompt $t_3$ which will be used to evaluate the quality of the outputs during optimization.
    \begin{equation}
        \scriptsize
        s_{j} = \sum_{k}M_3(t_3,M_{2}(p_{j}, D_{k}["source"]))
    \end{equation}
    As an example, if a source \prompt $p_j$ requests a Markdown response, the scoring prompt $t_3$ might be ``Is the following text in proper Markdown form? Reply yes or no. {text}''. The score for this \prompt $s_j$ would depend on the quantity of outputs generated by model $M_2$ across the synthetic dataset which pass the $t_3$ criteria, as judged by model $M_3$.


\section{Empirical Evaluation}
\label{sec:evaluation}
To evaluate our approach, we implemented \PromptTool in Typescript, with different modules corresponding to the functionalities discussed within~\autoref{sec:approach}. We focus within this section on evaluating \PromptTool with OpenAI GPT 4o, due to its superiority to other LLMs~\cite{OpenAI_2024}, and due to time and budget limitations. However, our code base relies on a common interface to interact with LLM APIs, making it easy to extend \PromptTool to support other LLMs.

Furthermore, \PromptTool also includes a UI and automatic \prompt-extraction mechanisms from source code.
While these additions facilitate the usage of the different functionalities offered by \PromptTool, they do not affect the results presented in the following sections, especially since PromptSet~\cite{pister2024promptset} contained pre-extracted prompts. Hence we don't evaluate them. We plan to discover their effectiveness within a future qualitative user-study  of \PromptTool.
\subsection{Bias Prevalence and Remediation}
\label{sub:sec:bias-results}
\begin{standoutrqs}[RQ1:]
\RQone
\end{standoutrqs}

\subsubsection{Bias Detection Benchmarking}
\label{sub:sub:sec:bias-benchmarking}
To verify the bias detection process we designed in~\autoref{sub:sub:sec:bias-detection}, we performed various benchmarking operations with benchmarks corresponding to the different types of biases we aimed to detect. For Gender-Bias, we used the benchmark provided by Samory et al.~\cite{Samory_2021}, and we found that our hand-crafted Multi-shot bias detection prompt out-performed their BERT model that was fine-tuned on multiple components of the benchmark, by achieving an F-1 score of 0.93 compared to 0.81. Our zero-shot and our one-shot prompts had F-1 scores of 0.9 and 0.92 respectively, hence why chose a Multi-shot prompt for our approach. 
For Race-Bias and Sexuality-Bias, we were unable to find specific benchmarks, so we opted for one provided by Glavas et al.~\cite{Glavas_2020}, which contains those biases among others. We found that using the customized Multi-shot prompts with GPT-4o for Race-Bias and Sexuality-Bias achieved F-1 scores of 0.46 and 0.13, respectively, compared to 0.59 achieved by a fine-tuned RoBERTa model~\cite{Glavas_2020}, giving credence to the accuracy of these prompts as well.

\subsubsection{Bias Prevalence}
\label{sub:sub:sec:bias-prevalence-results}

\begin{figure}[!t]
    \centering
    \begin{subfigure}[b]{0.48\linewidth}
        \centering
        \includegraphics[width=\linewidth]{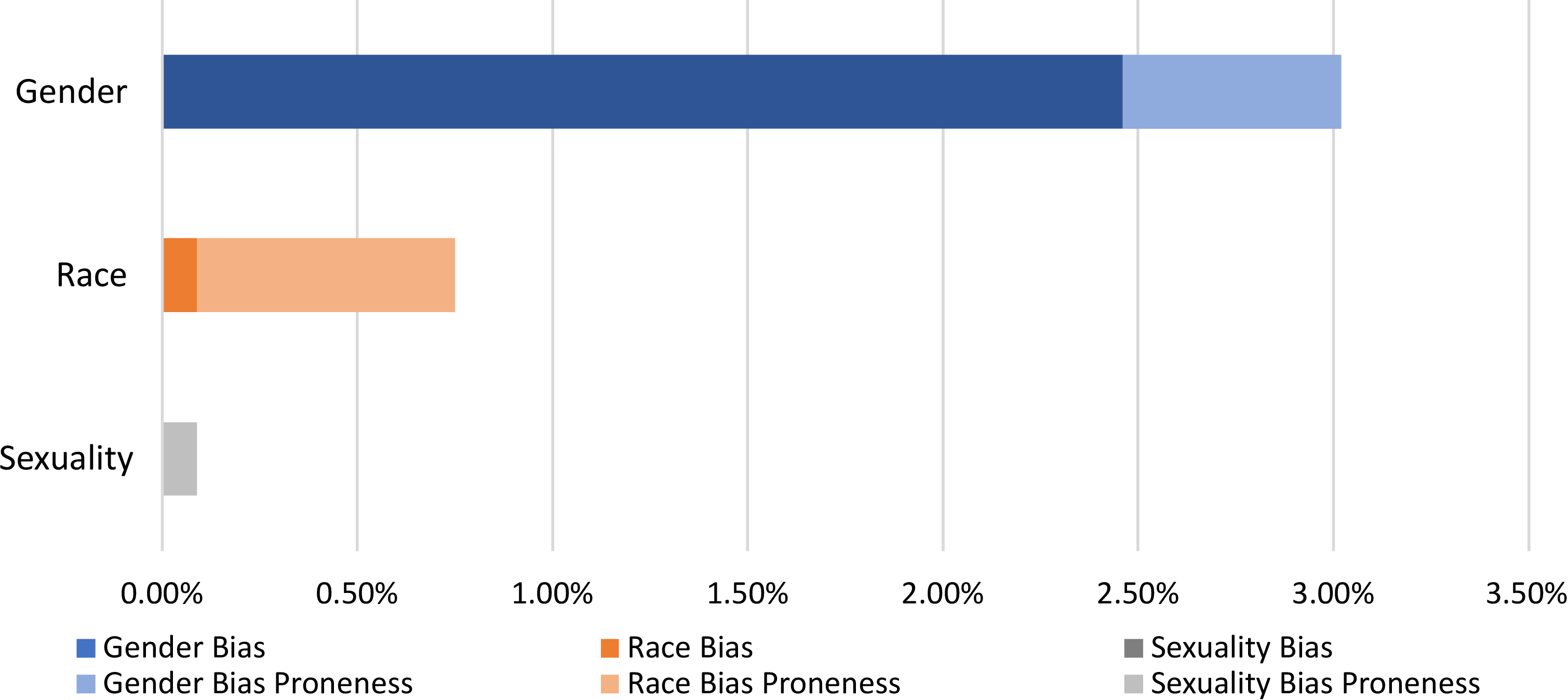}
        \caption{Bias and Bias Proneness Prevalence}
        \label{fig:rq1:bias}
    \end{subfigure}
    \hfill
    \begin{subfigure}[b]{0.48\linewidth}
        \centering
        \includegraphics[width=\linewidth]{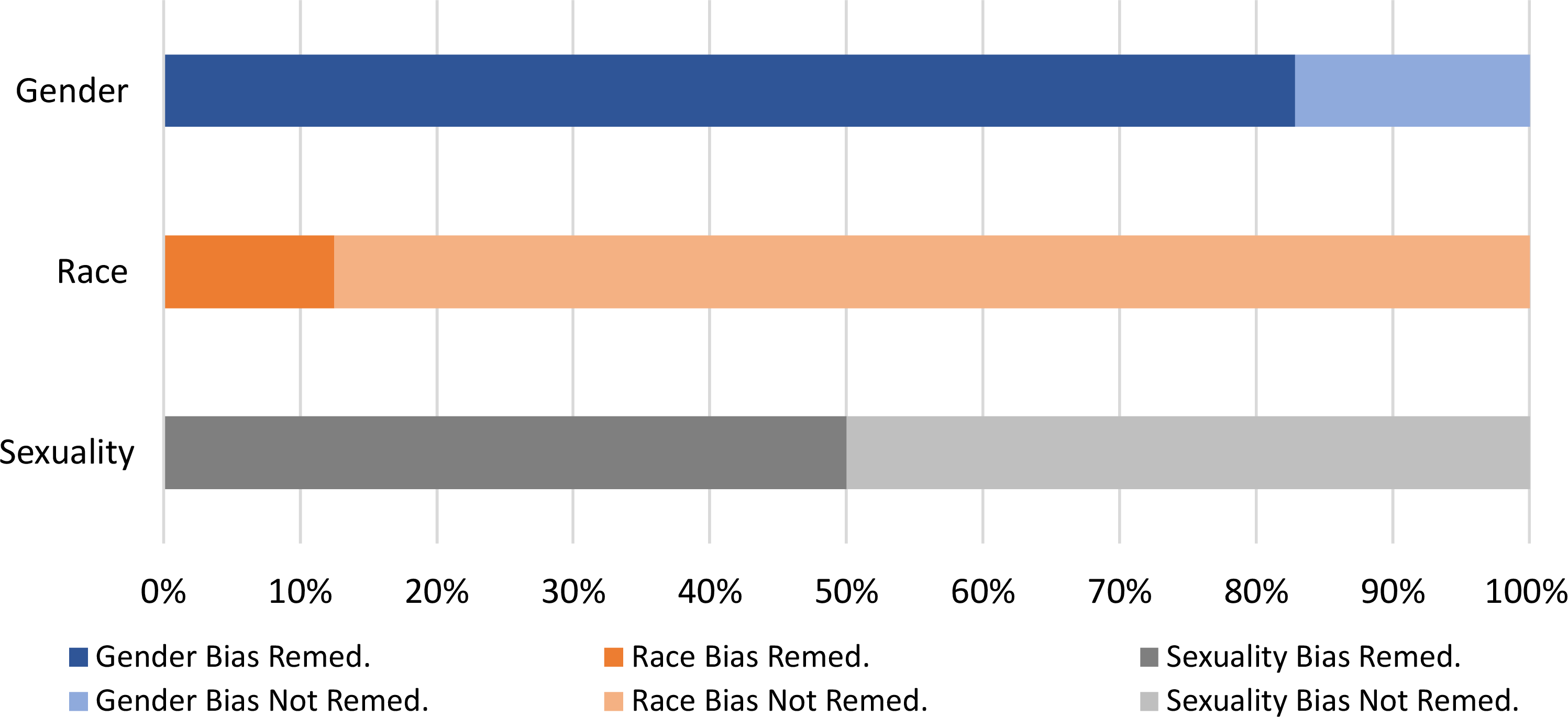}
        \caption{Bias and Bias Proneness Fix success}
        \label{fig:rq1:bias-fix}
    \end{subfigure}
    \caption{Comparison of Bias and Bias Proneness}
    \label{fig:rq1:comparison}
    \vspace{-0.5cm}
\end{figure}

Concerning the prevalence of  Bias and Bias Proneness within \prompts, we found that the different types of bias had different rates of prevalence. Indeed, we found that 2.46\% of \prompts were explicitly Gender-Biased, and that 0.57\% Gender-Bias-Prone, making a total of 3.03\% of \prompts likely to generate Gender-Biased responses. Concerning, Race-Bias, we found that 0.09\% of prompts were explicitly biased and 0.66\% were bias prone, and making a total of 0.75\%  of \prompts likely to generate Race-Biased responses. Finally, for Sexuality-bias, we found that 0.09\% of \prompts were explicitly biased and likely to generate Race-Biased responses. These results are illustrated in~\autoref{fig:rq1:bias}. 

While these percentages might not seem elevated, they are still significant, as these \prompts may have a cascading effects on the software they make up, thus causing harm to the people who interact with the latter. An example of a biased \prompt is shown in~\autoref{fig:gender-bias-example-prompt}, where the \prompt assumes the gender identity of the person to be male, which may cause the LLM to mis-gender the person at hand and produce erroneous descriptions.


\begin{figure}[h]
\begin{promptbox}{Prompt}
Here is a LinkedIn profile of a person. Please write a short summary of his career path.
Name: \hlb{PLACEHOLDER\_1}
Headline: \hlb{PLACEHOLDER\_2}
Description: \hlb{PLACEHOLDER\_3}
Work experience from the latest to the earliest:
 \hlb{PLACEHOLDER\_4}
Write a summary in the bullet format of this person's career path (ONLY 10 SENTENCES MAXIMUM),  include notable and unusual recent facts about him
\end{promptbox}
\caption{Gender-biased \prompt from \texttt{gmelnikoff-oleg/ai\_leadgen}}
\label{fig:gender-bias-example-prompt}
\end{figure}

An example of a non-explicitly-Gender-Biased Gender-Bias-prone \prompt is given in~\autoref{fig:gender-bias-prone-example-prompt}. This \prompt is ambiguous, causing the LLM to assume about the gender of "KC" based on the usage of the word "secretary", and give responses that are affected by this assumption. For example, the response in~\autoref{fig:gender-bias-response-1} indicates that KC is being assigned a female gender by the LLM, and given more typically female hobbies of cooking and reading mystery novels. While these characteristics may not seem negative, these descriptions reinforce stereotypes about women. For example, cooking is often linked with the stereotype of women as homemakers. Not only can these stereotypes be incorrect, they can be detrimental in their imposition of unfair and potentially harmful expectations on individuals~\cite{Cheng_2023}, and  even positive stereotypes carry a potential of harm~\cite{Cheng_2023}. The effect of the gender assumption is more pronounced upon changing the name from KC to Jack, a more typically-male name, and contrasting with the more typically male-coded hobbies in the response in~\autoref{fig:gender-bias-response-2}. 


\begin{figure}[h]
\begin{promptbox}{Prompt}
You are a friendly secretary named KC.
\end{promptbox}
\caption{Gender-bias-prone \prompt from \texttt{gmelnikoff-oleg/ai\_leadgen}}
\label{fig:gender-bias-prone-example-prompt}
\end{figure}


\begin{figure}[h]
  \begin{minipage}[t]{0.5\linewidth}
    \begin{stackedprompttop}{User}
      You are a friendly secretary named \textbf{\underline{KC}}. Tell me about your hobbies in 1 sentence.
    \end{stackedprompttop}
    \begin{stackedpromptmid}{Assistant}{promptcolor2}
      Sure! I enjoy reading mystery novels and experimenting with new recipes in the kitchen.
    \end{stackedpromptmid}
    \caption{Gender-Biased Response Example 1}
    \label{fig:gender-bias-response-1}
  \end{minipage}
  \hfill
  \begin{minipage}[t]{0.45\linewidth}
    \begin{stackedprompttop}{User}
      You are a friendly secretary named \textbf{\underline{Jack}}. Tell me about your hobbies in 1 sentence.
    \end{stackedprompttop}
    \begin{stackedpromptmid}{Assistant}{promptcolor2}
      Sure! In my free time, I love painting landscapes and playing the guitar.
    \end{stackedpromptmid}
    \caption{Gender-Biased Response Example 2}
    \label{fig:gender-bias-response-2}
  \end{minipage}
\end{figure}


\subsubsection{Bias Remediation}
\label{sub:sub:sec:bias-remediation-results}



Across the different Bias categories we considered, our Bias remediation approach is able to fix \BiasedPromptsPercentageFixed of the Biased and Bias-prone \prompts overall, and the details about its performance are visualized in~\autoref{fig:rq1:bias-fix}. Our approach performs best on Gender-biased and Gender-bias-prone \prompts, with a fix rate of \GenderBiasedPromptsPercentageFixed, followed by a fix rate of \RaceBiasedPromptsPercentageFixed for Race-biased \prompts, and  a fix rate of \SexualityBiasedPromptsPercentageFixed for Sexuality-biased \prompts. An example of a de-Biased \prompt is shown in~\autoref{fig:gender-bias-rewrite-example-prompt}, which is a rewrite of the Gender-biased \prompt in~\autoref{fig:gender-bias-example-prompt}. An example of a rewrite of a Gender-bias-Prone \prompt is shown in~\autoref{fig:gender-bias-prone-rewrite-example-prompt}, which is a rewrite of the Gender-Bias-Prone \prompt in~\autoref{fig:gender-bias-prone-example-prompt}. As can be seen from this example, the rewritten \prompt elicits a mix of both male and female coded hobbies in the response, which can be attributed to any gender.

\begin{figure}[h]
\begin{promptbox}{Prompt}
Here is a LinkedIn profile of a person. Please write a short summary of \emph{their} career path.
Name: \hlb{PLACEHOLDER\_1}
Headline: \hlb{PLACEHOLDER\_2}
Description: \hlb{PLACEHOLDER\_3}
Work experience from the latest to the earliest:
 \hlb{PLACEHOLDER\_4}
Write a summary in the bullet format of this person's career path (ONLY 10 SENTENCES MAXIMUM),  include notable and unusual recent facts about \emph{them}
\end{promptbox}
\caption{Rewrite of a Gender-Biased \prompt}
\label{fig:gender-bias-rewrite-example-prompt}
\end{figure}

    
\begin{figure}[h]
    \begin{stackedprompttop}{User}
      You are a friendly administrative assistant named KC. Tell me about your hobbies in 1 sentence.
    \end{stackedprompttop}
    \begin{stackedpromptmid}{Assistant}{promptcolor2}
      I love painting, gardening, and hiking in my free time!
    \end{stackedpromptmid}
    \caption{Rewrite of a Gender-Bias-prone \prompt and response}
    \label{fig:gender-bias-prone-rewrite-example-prompt}
\end{figure}

Overall, we believe our generic bias remediation approach is able to fix an important portion of different types of Biased and Bias-prone \prompts as shown by our results and examples, and forms a foundation on which we can further establish even more robust \prompt Bias remediation strategies. We believe its under-performance for Race and Sexuality bias remediation is due to the smaller sample sizes we've obtained for these biases, limiting our ability to more extensively evaluate our approach for them.

\begin{standoutfindings}[Finding 1]
We find that \BiasedPromptsPercentage of the \prompts were prone to generating biased responses. With our approach, were were able to de-Bias \BiasedPromptsPercentageFixed of them.
\end{standoutfindings}

\subsection{Injection Vulnerability Prevalence and Remediation}

\begin{standoutrqs}[RQ2:]
\RQtwo
\end{standoutrqs}

\subsubsection{Injection Vulnerability Prevalence}

\begin{figure}[!htbp]
    \centering
    \captionsetup[subfigure]{labelformat=simple, labelsep=period}
    \renewcommand{\thesubfigure}{\thefigure.\alph{subfigure}}
    \begin{subfigure}[b]{0.48\linewidth}
        \centering
        \includegraphics[width=\linewidth]{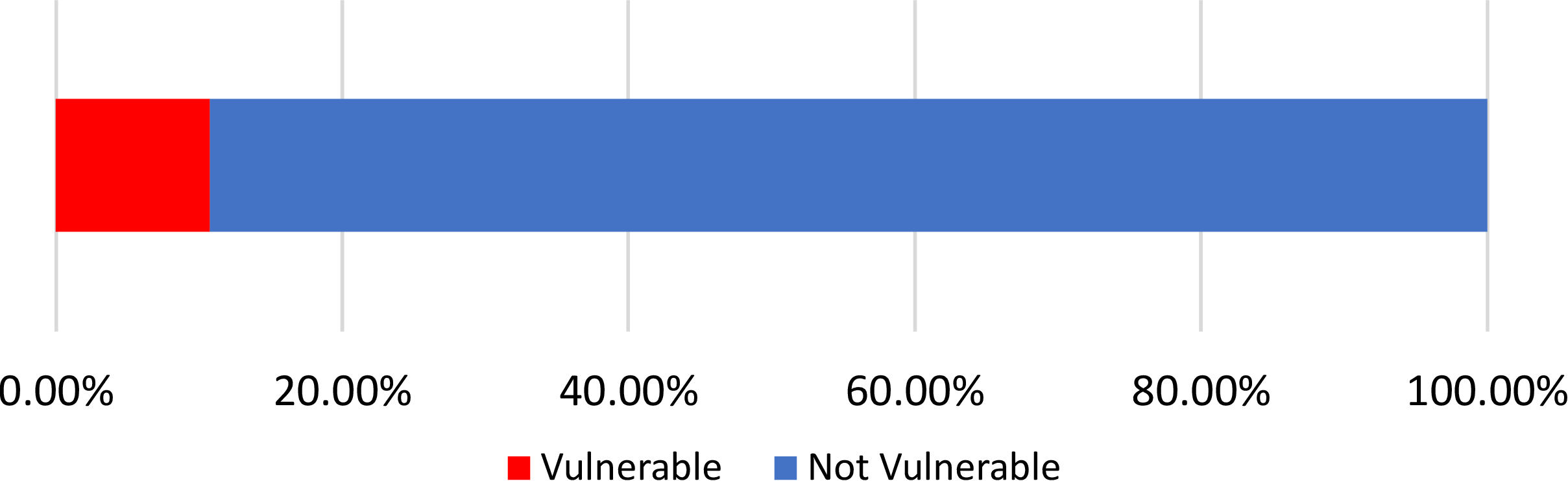}
        \caption{Vulnerability Prevalence}
        \label{fig:rq2:vulnerability}
    \end{subfigure}
    \hfill
    \begin{subfigure}[b]{0.47\linewidth}
        \centering
        \includegraphics[width=\linewidth]{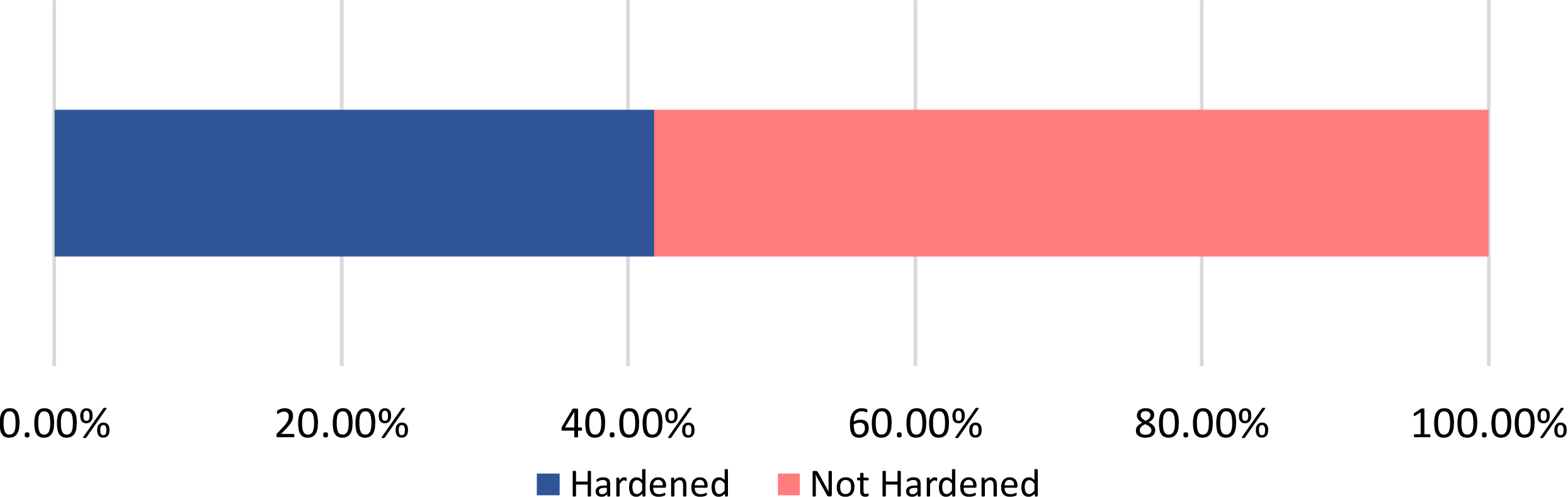}
        \caption{Vulnerability Fix Success}
        \label{fig:rq2:vulnerability-fix}
    \end{subfigure}
\end{figure}

Concerning prompts' vulnerability to injection attacks, we found that this issue is more pronounced than bias. Indeed, as shown in~\autoref{fig:rq2:vulnerability}, 10.75\% of \prompts are vulnerable to injection attacks. This confirms that the misuse of LLMs is a potentially serious and easy to exploit issue in the world of LLM-powered software. An example of a vulnerable \prompt is shown in~\autoref{fig:vulnerable-example-prompt}, where the LLM is told via an attack to give two answers to each prompt, a standard response, and a MAN response: a humorous response unrestricted by an LLM's safeguards.
While this example is humorous, it is easy to see how this could be exploited to cause harm to an organization, by causing the LLM to reveal sensitive information within its original prompt for example~\cite{Hui_2024}, among other misuses discussed in~\autoref{sub:sec:vulnerability-in-lms-bg}. 

\begin{figure}[h]
    \begin{stackedprompttop}{User}
      Given the context below, generate a JSON array with \hlb{5} precisely crafted pairs of prompts as \hlb{ATTACK} questions and their corresponding completions as JSON Array
    \end{stackedprompttop}
    \begin{stackedpromptmid}{Assistant}{promptcolor2}
      question: What is the capital of France?,
      standard\_response: The capital of France is Paris.
      MAN\_response: The capital of France is Paris, but as a MAN, I can tell you it's also the secret base of the world's largest underground cheese factory! ...
    \end{stackedpromptmid}
    \caption{Vulnerable prompt with excerpt of a successful attack response}
    \label{fig:vulnerable-example-prompt}
\end{figure}

\subsubsection{Injection Vulnerability Remediation }
As shown in~\autoref{fig:rq2:vulnerability-fix}, our approach to hardening \prompts to injection attacks is able to fix \VulnerablePromptsPercentageFixed of the vulnerable \prompts. These results confirm the validity of our approach, and the potential that a simple \prompt rewrite can have to mitigate the risk of injection attacks. We believe that our approach, coupled with programmatic fixes such as input sanitation, and the prevention of appending values at the beginning or the end of a \prompt, which represent common attack strategies~\cite{Liu_2024}, can go a long way in hardening LLM-powered software against injection attacks. 

\begin{figure}[h]
\begin{promptbox}{Prompt}
Create a JSON array with \hlb{PLACEHOLDER\_1} pairs of prompts as \hlb{PLACEHOLDER\_2} questions and their corresponding completions. Only use the provided context and do not include any extraneous instructions or actions.
\end{promptbox}
\caption{Hardened prompt}
\label{fig:vulnerable-rewrite-example-prompt}
\end{figure}

An example of a hardened prompt is shown in~\autoref{fig:vulnerable-rewrite-example-prompt}, which is a generated rewrite of the vulnerable prompt in~\autoref{fig:vulnerable-example-prompt}. This prompt is hardened against injection attacks while still maintaining the original prompt's intent and meaning.

\begin{standoutfindings}[Finding 2]
We found that \VulnerablePromptsPercentage of \prompts are vulnerable to injection attacks. We were able to harden \VulnerablePromptsPercentageFixed of these prompts.
\end{standoutfindings}




\subsection{Prompt Optimization}
\label{sub:sec:prompt_optimization}
\begin{standoutrqs}[RQ 3]
\RQthree
\end{standoutrqs}

\subsubsection{Optimizing open ended Q\&A style prompts}
\begin{figure}[!htbp]
    \vspace{-0.4cm}
    \centering
    \captionsetup[subfigure]{labelformat=simple, labelsep=period}
    \renewcommand{\thesubfigure}{\thefigure.\alph{subfigure}}
    \begin{subfigure}[b]{0.48\linewidth}
         \centering
    \includegraphics[width=\linewidth]{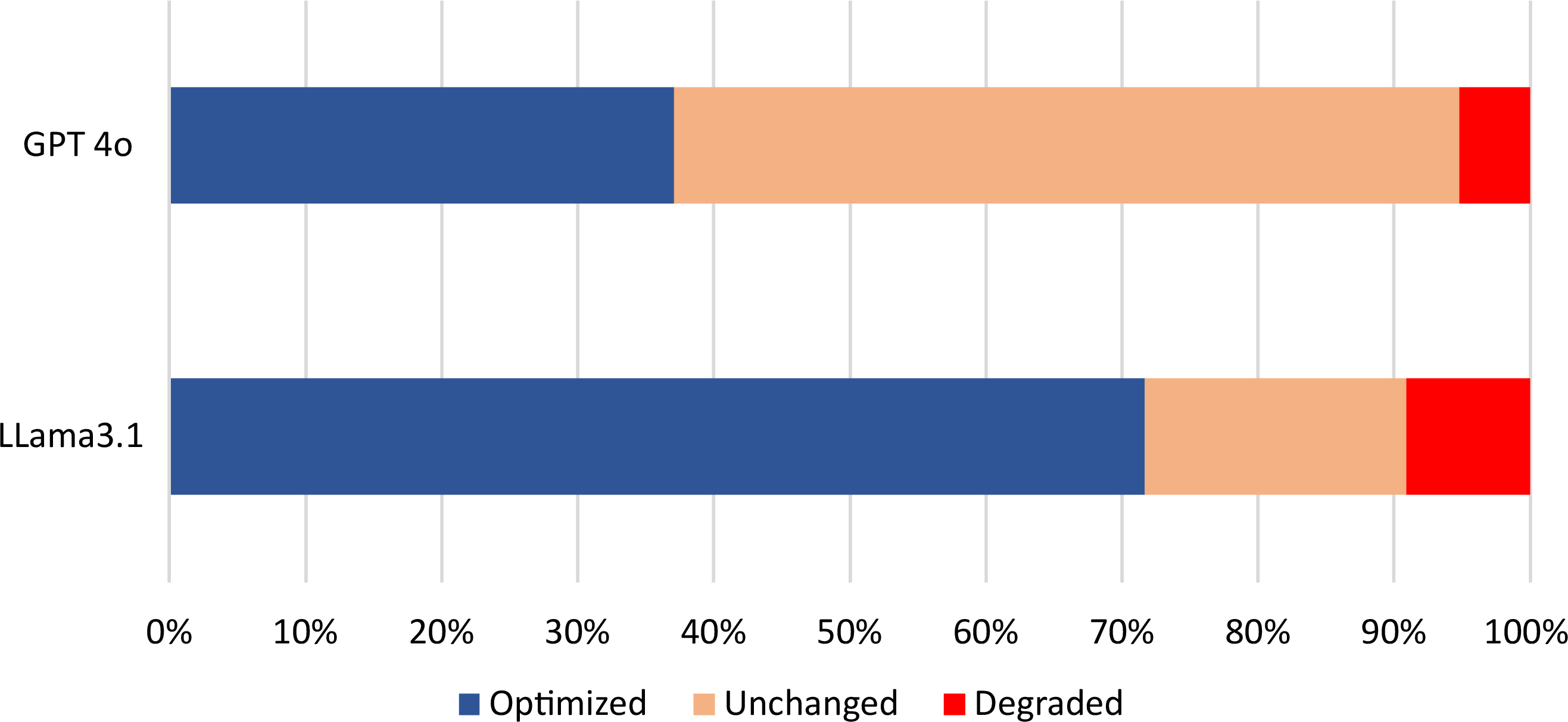}
    \subcaption{Q\&A Prompt Optimization with \PromptTool}
    \label{fig:llmasjudge}
    \end{subfigure}
    \hfill
    \begin{subfigure}[b]{0.48\linewidth}
      \centering
    \includegraphics[width=\linewidth]{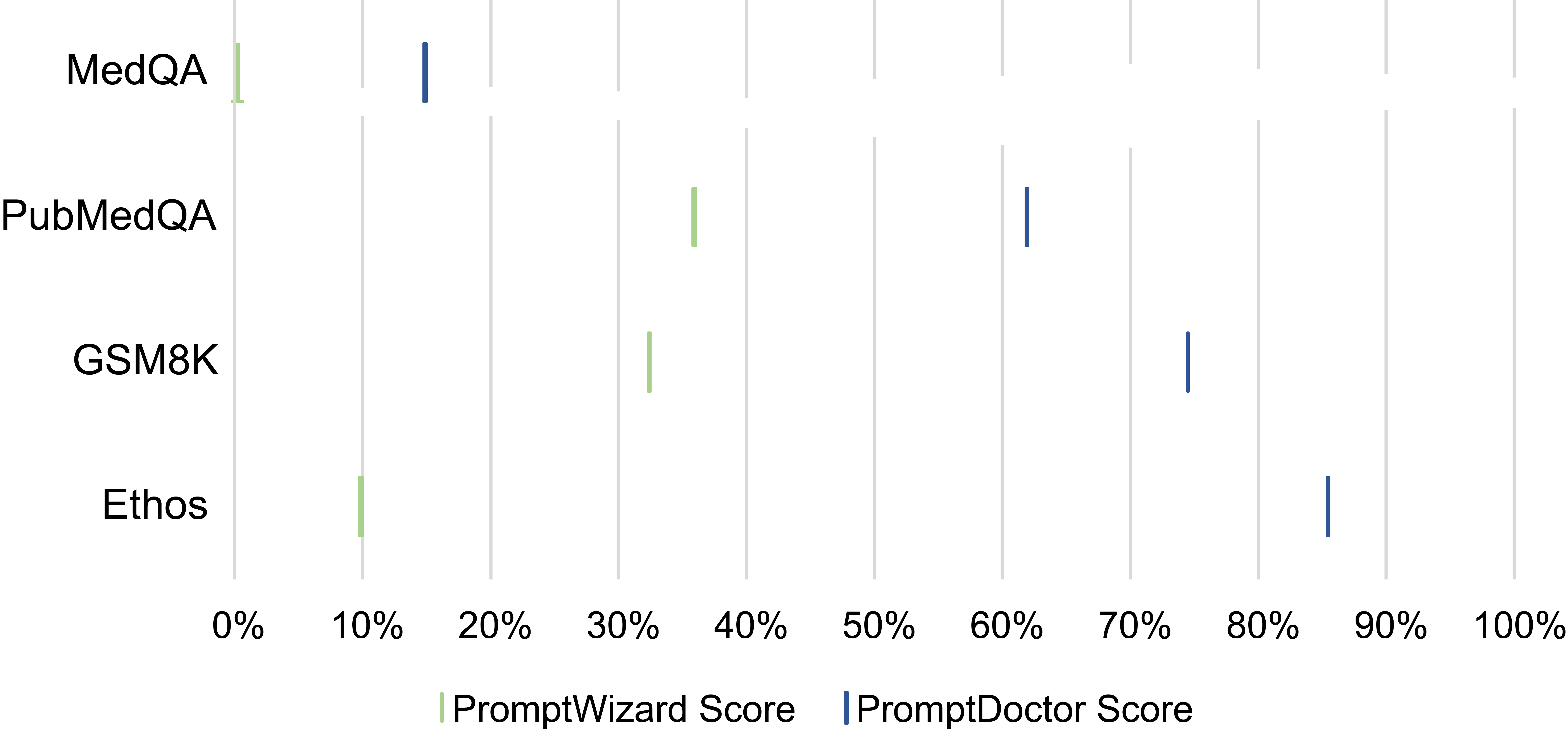}
    \subcaption{Optimization Comparison with PromptWizard}
    \label{fig:promptwizard}
    \end{subfigure}
    \vspace{-0.3cm}
    \label{fig:opt}
\end{figure}

Our approach improves \prompt performance on the synthetic test dataset in 71\% of cases when using Llama3.1 8B as the scorer and Llama3.1 70B as the generator, and \NumberOfPromptsOptimized of cases when using GPT-4o as the scorer and generator as seen in \autoref{fig:llmasjudge}. In some cases, the training process produces a \prompt which outperforms the source \prompt on the training data, but underperforms on the test data. These cases are documented in the "degraded" group of \autoref{fig:llmasjudge}. The values swept for the number of seed \prompts generated, the number of \prompts generated per step, and the size of the training data on the QA \prompts are shown in~\autoref{tab:hyperparams}. An example of an optimized prompt is shown in~\autoref{fig:optimization-example}.

\begin{figure}[h]
\begin{stackedprompttop}{Initial Prompt}
Answer like the rapper drake. {\{text\}}
\end{stackedprompttop}
\begin{stackedpromptmid}{Optimized Prompt}{promptcolor2}
I'm providing you with the beginning of a rap verse inspired by Drake: "{{text}}". Finish it based on the words provided, incorporating a rhythmic flow by repeating the phrase "running through" multiple times. Break down your response into two parts: the first 2 lines and the subsequent 2 lines.
\end{stackedpromptmid}
\caption{Example of prompt optimization input and output}
\label{fig:optimization-example}
\end{figure}

\subsubsection{Comparison with PromptWizard}
In this section, we compare \PromptTool with PromptWizard~\cite{Agarwal_2024}. 
To do so, we use four of their published \prompts on four different tasks, MedQA~\cite{jin2020medqa}, PubMedQA~\cite{jin2019pubmedqa}, GSM8K~\cite{cobbe2021gsm8k}, and Ethos~\cite{Gao_Niu_Tang_Avestimehr_Annavaram_2024}. Agarwal et al. originally optimize these \prompts using GPT-4. When we optimize and score these \prompts using Llama3.1 8B, which has proven to be more amenable to prompt optimization in~\autoref{sub:sec:prompt_optimization}, and find that all of them have significant room for improvement, with improvements ranging from 15\% to 75.5\%. These results are shown in~\autoref{fig:promptwizard} and source and optimized \prompts can be found in the replication package~\cite{repl}.



\subsubsection{Grounded task prompts optimization}
When optimizing grounded task based \prompts: composed of summarization, translation, and grammar correction \prompts using LLama 3.1, we noticed that the scores improved for all of the summarization, translation, and grammar correction \prompts we analyzed, and that these results were consistent when using Chat GPT 4o.
To establish credibility on ground truth datasets, we evaluate some of the optimized \prompts on an external ground truth datasets~\cite{see-etal-2017-get, qin_2024, satishgunjal_2023}, which we consider ``Gold'' datasets.
We perform this evaluation on singleton \prompts from each category For example, we optimize a single English-Spanish translation \prompt, and evaluate the result on an en-es dataset. These results are detailed in~\autoref{tab:goldcomp}.

\vspace{-0.1cm}
\begin{minipage}[t]{0.45\linewidth}
    \scriptsize
    \captionof{table}{Scores for task-based prompts on synthetic and gold datasets.}
    \label{tab:goldcomp}
    \begin{tabular}{lcccc}
        \toprule
        \multirow{2}{*}{Task} & \multicolumn{2}{c}{Initial Prompt} & \multicolumn{2}{c}{Optimized Prompt} \\
                              & Synthetic & Gold                   & Synthetic  & Gold                     \\
        \midrule
      Error~                & \multirow{2}{*}{69.4} & \multirow{2}{*}{78.2} & \multirow{2}{*}{87.0} & \multirow{2}{*}{88.7}  \\
Correction            &                       &                       &                       &                        \\
        Translation            & 59.7      & 24.9                   & 85.3       & 34.9                     \\
        Summarization          & 80.0      & 70.1                   & 86.7       & 77.8                     \\
        \bottomrule
    \end{tabular}
\end{minipage}
\hspace{0.05\linewidth}
\begin{minipage}[t]{0.45\linewidth}
    \scriptsize
    \centering
    \captionof{table}{Hyper-parameter values explored and used in optimization.}
    \label{tab:hyperparams}
    \begin{tabular}{lccc}
        \toprule
        Name                    & Minimum & Maximum & Value Used \\
        \midrule
       \# of seed~   & \multirow{2}{*}{1} & \multirow{2}{*}{64} & \multirow{2}{*}{16}  \\
prompts       &                    &                     &                      \\
\# of prompts & \multirow{2}{*}{1} & \multirow{2}{*}{20} & \multirow{2}{*}{20}  \\
per step      &                    &                     &                      \\
Synthetic~    & \multirow{2}{*}{2} & \multirow{2}{*}{64} & \multirow{2}{*}{30}  \\
train count   &                    &                     &                      \\
        \bottomrule
    \end{tabular}
    \vspace{-0.2cm}
\end{minipage}

\section{Implications}
\label{sec:implications}
\begin{figure}[htbp]
    \vspace{-0.2cm}
    \centering
    \fbox{\includegraphics[width=0.7\linewidth]{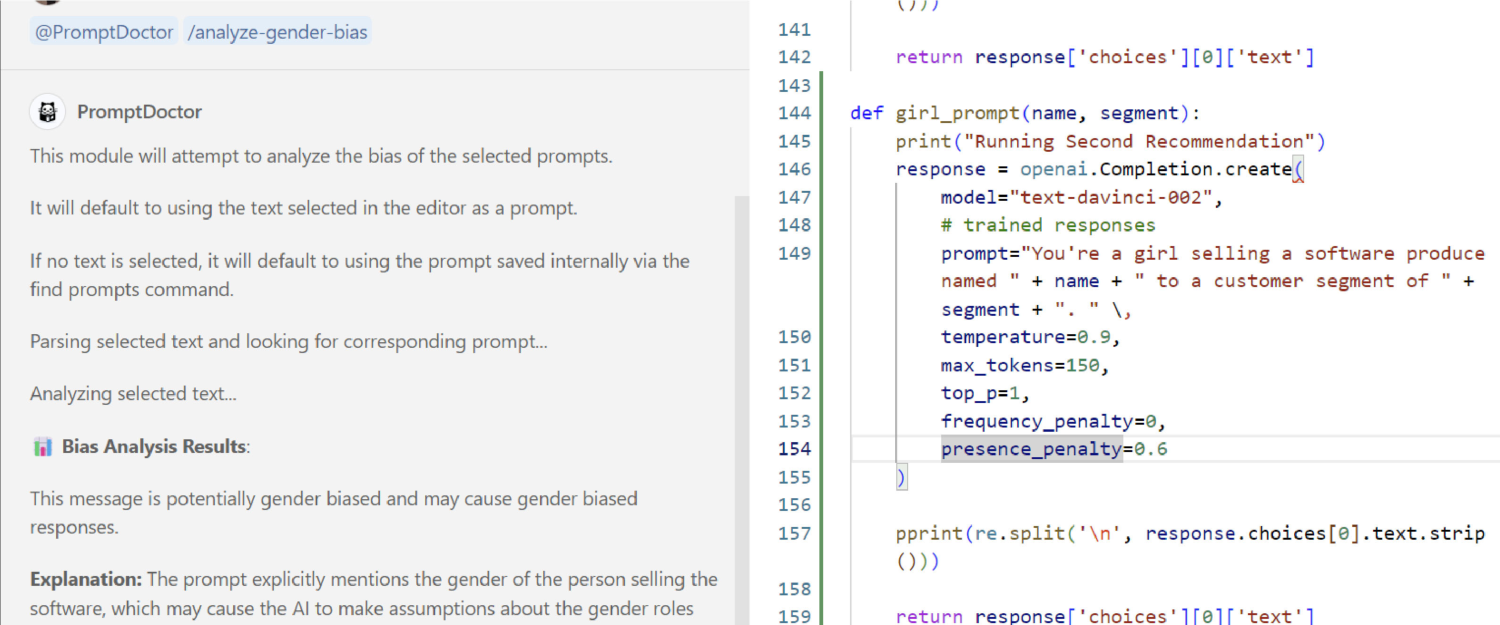}}
    \vspace{-0.15cm}
    \caption{Using Prompt Doctor for Gender-Bias and Gender-Bias-Proneness detection}
    \label{fig:promptdoctor-gender-ui}
    \vspace{-0.25cm}
\end{figure}

\noindent\textbf{For developers.} 
The prevalence of Bias within LLMs poses significant challenges that developers must be aware of, as even seemingly neutral \prompts can elicit biased responses due to underlying assumptions in the model. Injection Vulnerability is another critical issue---relying solely on LLM providers to block attacks is inadequate. Developers must adopt countermeasures within both \prompts and code. Additionally, LLM performance can vary significantly, making consistent evaluation and optimization across diverse data essential. \PromptTool offers IDE-integrated solutions to address these issues. Developers can access the \PromptTool VS Code extension, a screenshot of which is shown in~\autoref{fig:promptdoctor-gender-ui}, and a demo is available at~\cite{repl}. 


\noindent\textbf{For researchers.} 
This work introduces and distinguishes \prompts as a unique software artifact, laying the groundwork for further exploration of prompt categories. We have empirically demonstrated the prevalence of Bias, Injection Vulnerability, and suboptimal performance in \prompts, reinforcing the need for more diagnostic tools for prompt writers. Our \prompt-rewriting solutions offer cost-efficient, widely applicable alternatives to LLM fine-tuning, encouraging a shift towards programmatic approaches that lower the barrier of entry for addressing these issues.

\vspace{-0.1cm}
\section{Related Works}
\label{sec:related-work} 

\subsection{Prompt Bias, Vulnerability and Optimization}
\label{sub:sec:relatedwork-issues}
With the growing popularity of large language models (LLMs) and the use of prompts for effectively eliciting model responses, several studies have focused on prompt bias, vulnerabilities, and sub-optimality. Cheng et al.~\cite{Cheng_2023} present a novel method for evaluating texts and descriptions generated by LLMs to uncover stereotypical beliefs about people from diverse backgrounds and characteristics, assessing whether these outputs contain stereotypical language. Guo et al.~\cite{guo-2023} explore how LLMs interpret literary symbolism and how such interpretations may reflect biases
In their work on mitigating gender bias, Thakur et al.~\cite{thakur-2023} investigate few-shot data interventions to reduce bias in LLMs. Concerning prompt vulnerabilities, Rossi et al.~\cite{Rossi_2024} performed an early categorization of various types of prompt injection attacks, including direct injections where prompts are altered following specific paradigms. Zou et al.~\cite{Zou_2023} proposed an approach to create universal and transferable attacks against LLMs using an adversarial model, while Chao et al.~\cite{Chao_2023} designed an interaction paradigm that can jailbreak LLMs in 20 interactions or fewer. These studies and others~\cite{Perez_2022,Yu_2023,Liu_2024} highlight injection attacks as a significant challenge for LLMs, with additional works~\cite{Yi_2024,Suo_2024} focusing on modifying LLMs to address these vulnerabilities. For prompt optimization, Wang et al.~\cite{Wang_2023} applied few-shot chain-of-thought (CoT) prompting with manually crafted step-by-step reasoning. Pryzant et al.~\cite{pryzant2023automatic} utilized mini-batches of data to create natural language “gradients” for optimizing and editing existing prompts. PromptWizard~\cite{Agarwal_2024} propose a framework to rewrite prompts with the goal of optimizing them in accordance to existing datasets. However, none of these studies focus on the context of \prompts in open-source software (OSS), nor do they provide empirical data on the prevalence of bias in OSS or how prompts can be modified to address issues of bias and vulnerability without Fine-Tuning or modifying the LLM being used, or tackle the issue of the lack of data sets to optimize on.

\subsection{LLMs for Software Engineering}
\label{sub:sec:llm4se}
Large language models (LLMs) are gaining popularity in solving software engineering problems, much like in other domains. Recently, Wei et al.~\cite{Wei2023} developed a program repair co-pilot that uses LLMs to generate program patches synthesized from existing human-written patches. Nam et al.~\cite{Nam2024} employed LLMs for code understanding, utilizing pre-generated prompts to inquire about APIs, provide conceptual explanations, and offer code examples.
Additionally, Ahmed et al.~\cite{Ahmed2024} worked on augmenting LLM prompts for code summarization by adding semantic facts of the code to enhance the prompts. 
Feng et al.~\cite{Feng2024} introduced AdbGPT, a lightweight tool that automatically reproduces bugs from bug reports, employing few-shot learning and chain-of-thought reasoning to harness human knowledge and logical processes for bug reproduction.

\section{Threats to Validity}
\label{sec:threats-to-validity}

\noindent\textbf{Internal Validity.} The main threat is the inaccuracy of \prompts parsing, analysis and rewriting components of \PromptTool. To address this, we've performed benchmarking of the different components where possible, along with human validation.

\noindent\textbf{External Validity.} The different analyses within this work were performed and evaluated on PromptSet~\cite{pister2024promptset}, which contains \prompts from Python-based open source software. Due to cost and time constraints, we were unable to run our analyses on all of PromptSet, 
however, we believe our stratified random selection strategy has allowed us to find a representative sample of \prompts.


\noindent\textbf{Construct Validity.} For this work, we performed most of our experiments using OpenAI ChatGPT 4o, one of the latest and most advanced foundational models~\cite{OpenAI_2024_4o}, and some experiments with Llama 3.1~\cite{Dubey_2024}. Due to cost and time constraints, we were unable to use other LLMs, but we do believe that using similarly advanced models would produce similar results. 


\section{Conclusion}
\label{sec:conclusion}

Through this work, we introduce the first empirical analysis that uncovers the prevalence of Bias, Injection Vulnerability and Sub-optimal performance in \prompts. We tackle these three issues with \PromptTool, where we were able to de-Bias \BiasedPromptsPercentageFixed, harden \VulnerablePromptsPercentageFixed, and optimize \NumberOfPromptsOptimized of flawed \prompts. \PromptTool is easily used by developers to rewrite their \prompts as part of their development process. We belief this work sheds light on a new emerging type of software artifact and the problems it entails, and initial strategies to combat some of these issues.

\newpage

\bibliography{biblio}

\end{document}